\documentclass[11pt]{article}
\usepackage{graphicx}
\usepackage{cite}

\RequirePackage{xspace}

\def\Bbar    {\kern 0.18em\overline{\kern -0.18em B}{}\xspace}
\def\Dbar    {\kern 0.18em\overline{\kern -0.18em D}{}\xspace}
\def\sbar    {\kern 0.18em\overline{\kern -0.18em s}{}\xspace}
\def\Kbar    {\kern 0.18em\overline{\kern -0.18em K}{}\xspace}
\def\cbar    {\kern 0.18em\overline{\kern -0.18em c}{}\xspace}

\def\BBbar{\mbox{$B\overline {B}\ $}}
\def\Bzb     {\ensuremath{\Bbar^0}\xspace}
\def\K0S         {\ensuremath{K^0_S}\xspace}
\def\CP                {\ensuremath{C\!P}\xspace}
\def\CM                {\ensuremath{C\!M}\xspace}
\def\ML                {\ensuremath{M\!L}\xspace}
\def\LR                {\ensuremath{L\!R}\xspace}
\def\CKM                {\ensuremath{C\!K\!M}\xspace}

\def\SM                {\ensuremath{S\!M}\xspace}
\def\NP                {\ensuremath{N\!P}\xspace}
\def\MSSM                {\ensuremath{M\!S\!S\!M}\xspace}

\def\UL                {\ensuremath{U\!L}\xspace}
\def\CL                {\ensuremath{C\!L}\xspace}

\def\Dstarz  {\ensuremath{D^{*0}}\xspace}
\def\Dstarp  {\ensuremath{D^{*+}}\xspace}
\def\Dstar   {\ensuremath{D^*}\xspace}
\def\Dstarpm   {\ensuremath{D^{*\pm} } \xspace}

\def\RDx{\ensuremath{{\cal R}(D^{(*)})} \xspace}
\def\ds{\ensuremath{D^{(*)}}\xspace}

\def\taum       {\ensuremath{\tau^-}\xspace}
\def\nub        {\ensuremath{\overline{\nu}}\xspace}
\def\nutb       {\ensuremath{\nub_\tau}\xspace}

\def\Dz      {\ensuremath{D^0}\xspace}
\def\Dp      {\ensuremath{D^+}\xspace}
\def\Dm      {\ensuremath{D^-}\xspace}
\def\Dst      {\ensuremath{D^*}\xspace}
\def\Dzb     {\ensuremath{\Dbar^0}\xspace}
\def\epem  {\ensuremath{e^+\!e^-}\xspace}

\def\Kz      {\ensuremath{K^0}\xspace}
\def\Kzb      {\ensuremath{\Kbar^0}\xspace}

\def\K0S      {\ensuremath{K^0_S}\xspace}

\def\Nubar    {\kern 0.18em\overline{\kern -0.18em \nu}{}\xspace}
\def\ra                 {\ensuremath{\rightarrow}\xspace}
\def\SM                {\ensuremath{S\!M}\xspace}
\def\CL                {\ensuremath{C\!L}\xspace}
\def\NP                {\ensuremath{N\!P}\xspace}
\def\BDT                {\ensuremath{B\!D\!T}\xspace}
\def\FCNC          {\ensuremath{F\!C\!N\!C}\xspace}
\def\calA{{\ensuremath{\cal A}}\xspace}
\def\calAbar    {\kern 0.18em\overline{\kern -0.18em \calA}{}\xspace}

\def\babar{{\em B}{\footnotesize\em A}{\em B}{\footnotesize\em AR}}

\def\calR{{\ensuremath{\cal R}(D)}\xspace}
\def\calRst{{\ensuremath{\cal R}(\Dst)}\xspace}

\def\AntiBDtauNu{{\ensuremath{\Bbar \ra D\tau\nu_{\tau}}\xspace}}
\def\AntiBDlNu{{\ensuremath{\Bbar \ra D l \nu_{l}}\xspace}}

\def\AntiBDsttauNu{{\ensuremath{\Bbar \ra \Dst \tau \nu_{\tau}}\xspace}}
\def\AntiBDstlNu{{\ensuremath{\Bbar \ra \Dst l \nu_{l}}\xspace}}

\def\bbar  {\mbox{$\bar{b} $}}
\def\dbar  {\mbox{$\bar{d} $}}

\providecommand{\UfourS}{\mbox{$\Upsilon(4S)$}}
\newcommand{\etal}{{\em et al.}}
\newcommand{\jprlBase}       {Phys.\ Rev.\ Lett.\xspace}
\newcommand{\jprl}      [1]  {\jprlBase\ {\bf #1}}
\newcommand{\rmpBase}       {Rev.\ Mod.\ Phys.\xspace}
\newcommand{\rmp}      [1]  {\rmpBase\ {\bf #1}}
\newcommand{\jprBase}        {Phys.\ Rev.\xspace}
\newcommand{\jprd}      [1]  {\jprBase\ D~{\bf #1}}
\newcommand{\jplBase}        {Phys.\ Lett.\xspace}
\newcommand{\plb}       [1]  {\jplBase\ B~{\bf #1}}
\newcommand{\jnpBase}        {Nucl.\ Phys.\xspace}
\newcommand{\npb}       [1]  {\jnpBase\ B~{\bf #1}}

\newcommand{\epjBase}       {Eur.\ Phys.\ J.\xspace}
\newcommand{\epjc}      [1]  {\epjBase\ C~{\bf #1}}
\newcommand{\jhepBase}       {JHEP\xspace}
\newcommand{\jhep}      [1]  {\jhepBase\ ~{\bf #1}}
\newcommand{\progtpBase}       {Prog.\ Theor.\ Phys.\xspace}
\newcommand{\progtp}      [1]  {\progtpBase\ ~{\bf #1}}
\newcommand{\jpBase}       {J.\ Phys.\xspace}
\newcommand{\jp}      [1]  {\jpBase\ G ~{\bf #1}}
\newcommand{\zpcBase}       {Z.\ Phys.\xspace}
\newcommand{\zpc}      [1]  {\zpcBase\ C ~{\bf #1}}
\newcommand{\nppsBase}        {Nucl.\ Phys.\ Proc.\ Suppl.\xspace}
\newcommand{\npps}       [1]  {\nppsBase\ ~{\bf #1}}

\newcommand{\mev}{\ensuremath{{\mathrm{\,Me\kern -0.1em V}}}\xspace}
\newcommand{\mevcc}{\ensuremath{{\mathrm{\,Me\kern -0.1em V\!/}c^2}}\xspace}
\newcommand{\mevc}{\ensuremath{{\mathrm{\,Me\kern -0.1em V\!/}c}}\xspace}
\newcommand{\gev}{\ensuremath{{\mathrm{\,Ge\kern -0.1em V}}}\xspace}
\newcommand{\gevcc}{\ensuremath{{\mathrm{\,Ge\kern -0.1em V\!/}c^2}}\xspace}
\newcommand{\gevc}{\ensuremath{{\mathrm{\,Ge\kern -0.1em V\!/}c}}\xspace}
\newcommand{\gevccq}{\ensuremath{{\mathrm{\,Ge\kern -0.1em V^2\!/}c^4}}\xspace}

\begin{document}
{\pagestyle{empty}

\begin{flushright}
\babar-PROC-11-062
\end{flushright}

\par\vskip 4cm

\begin{center}
\Large \bf  Recent \CKM and  and \CP Results from    \babar
\end{center}
\bigskip

}
\begin{center}
\large 
F. Palombo\\
Universit\`a degli Studi di Milano, Dipartimento di Fisica and INFN, I-20133
Milano, Italy \\
{\it on behalf of the \babar\  Collaboration}
\end{center}
\bigskip \bigskip

\begin{center}
\large \bf Abstract 
\end{center}
We present   recent  results of $B$ and charm decays from the \babar\ experiment. These results  include searches for rare or forbidden  charm decays, 
measurements of $|V_{ub}|$ from inclusive $\Bbar \ra X_u l \Nubar $ decays , observation of the    semileptonic $\Bbar \ra D^{(*)} \tau^-  \Nubar_{\tau}$ decays, direct \CP\ violation asymmetry in $B \ra X_{s+d}\gamma$ and in $D^+\ra K^0_S \pi^+$, and T-violation in $D^+_{(s)} \ra K^+ \K0S  \pi^+ \pi^-$. These studies are based on the final 
dataset collected  by \babar\   at the PEP-II   B factory at SLAC in the period 1999-2008.

\vfill
\newpage

\section{Introduction}

We present   some recent  results  concerning $B$ and charm decays from the \babar\ experiment. These results  include searches for rare or forbidden  charm decays, 
measurements of $|V_{ub}|$ from inclusive $\Bbar \ra X_u l \Nubar $ decays, observation of the    semileptonic $\Bbar  \ra D^{(*)}  \tau^- \Nubar_{\tau}$ decays, direct \CP\ violation asymmetry in $B\ra X_{s+d}\gamma$ and in $D^+\ra K^0_S \pi^+$, and T-violation in $D^+_{(s)} \ra K^+ \K0S  \pi^+ \pi^-$. 
The aim of these measurements is to test the Standard Model (\SM) mechanism of \CP violation and to search for possible contributions from  New Physics (\NP) beyond the \SM.  

 Important areas of search for  \NP\   are  processes which are expected at low level in \SM and which could be enhanced by \NP. 
This is the case for example of the measurement of branching fraction  of many rare decays , in particular  Flavor Changing Neutral Currents (\FCNC), and the measurement of \CP violating asymmetries. Such  measurements  are sensitive to \NP scenarios  and have been already  very powerful in constraining the parameter space of \NP\ models. 

The studies presented in this talk are based on the final (or almost final)  dataset collected  by \babar\  in the period 1999-2008 at the PEP-II B factory  at SLAC.

\section{\Dz  $\ra \gamma \gamma $  and \Dz $ \ra \pi^0 \pi^0$}
\FCNC\ decays, forbidden at tree level in \SM \cite{GIM}  but  allowed  at higher order,  have been already observed  in $K$ and $B$  meson systems \cite{TH}.  In the  charm sector a  \FCNC\  process  has an additional suppression   due to  the GIM mechanism \cite{GIM}.  
Thus far no charm \FCNC decay process has been observed. Interest in  \FCNC\  processes in the charm sector increased with the recent measurements  of \Dz\ - \Dzb mixing \cite{DDmixing}. Source of this mixing in fact can come from \NP, enhancing  the branching fraction  of \FCNC\  with respect to SM calculation \cite{Burdman, Prelo}. 

Calculations in the framework of vector meson dominance \cite{Burdman} and  of  heavy quark effective theory combined with chiral perturbation theory  \cite{Fajfer}   predict for   the  $D^0 \ra \gamma \gamma$ decay  the dominance of long-range effects
and  a branching fraction   of    ($3.5^{+4.0}_{-2.6} ) \times 10^{-8}$  and  $(1.0 \pm 0.5)\times 10^{-8}$, respectively. These branching fraction  estimates are orders of magnitude below the sensitivity of current experiments. But gluino exchange in Minimal Supersymmetric Standard Model (\MSSM)   can enhance the \SM \  branching fraction   up to  a factor 200 \cite{Prelo} (so within \babar\ sensitivity).  

CLEO    has measured  an  Upper Limit (\UL) of  $2.9 \times 10^{-5}$ at 90\%  Confidence Level (\CL)   for the branching fraction  of the decay $D^0 \ra \gamma \gamma$ \cite{CLEO} and a branching fraction  of  $(8.1 \pm 0.5)\times 10^{-4}$ for  the decay mode $\Dz \ra \pi^0 \pi^0 $ \cite{Mendez}. 

Recently \babar\  has studied  the decays $D^0 \ra \gamma \gamma$  and $\Dz \ra \pi^0 \pi^0$ with  a dataset  corresponding to an integrated luminosity of $470.5$ $ fb^{-1}$. \Dz  is reconstructed 
with a   $D^{*+} $  tag  in the decay $D^{*+} \ra \Dz \pi^+$ \cite{CC}, with the charge of the soft  pion from  $D^{*+}$ indicating the initial flavor of the \Dz.  This  tag suppresses the dominant combinatoric background.  To avoid uncertainties in the number of $D^{*+}$, the branching fractions of the decay modes $D^0 \ra \gamma \gamma$ and $D^0 \ra \pi^0 \pi^0$ are measured relative to the reference  decay mode $D^0 \ra \K0S  \pi^0$.  This mode has a large  and  precisely measured  branching fraction  ($1.22 \pm  0.05)  \times 10^{-2}$\cite{Nakamura}.
 
\begin{figure}[htb]
\begin{center}
\includegraphics[scale=0.35] {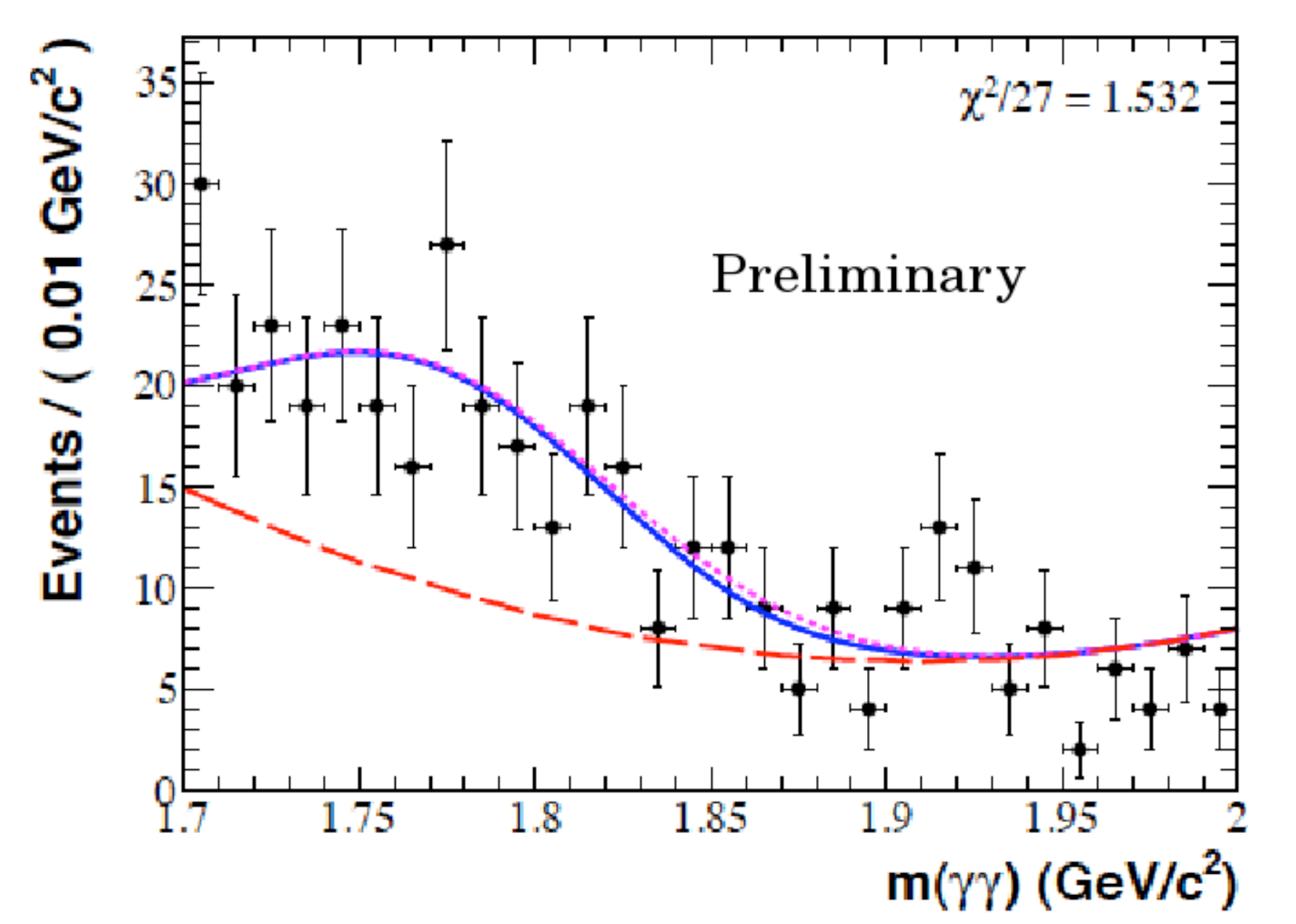} 
\hspace{1.5cm}
\includegraphics[scale=0.35] {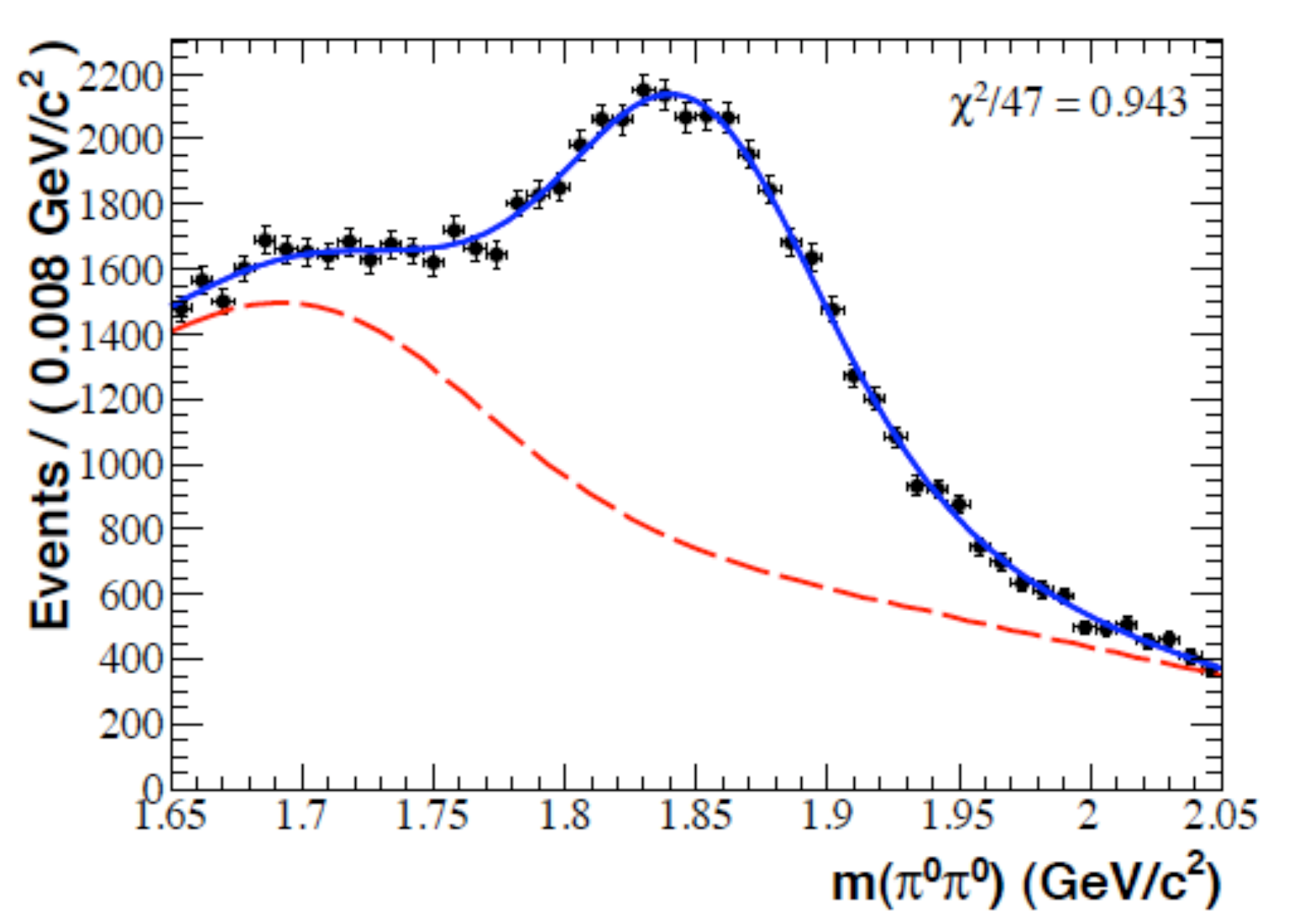} 
\caption{ Fit results to $\Dz \ra \gamma \gamma $ (on the left)  showing  data (points), the combinatoric background (long-dashed red),  the combinatoric background plus the $\Dz \ra \gamma \gamma$ background (short-dashed magenta) and and the total fit function including signal and the two backgrounds (solid blue);  fit results 
 to $\Dz \ra \pi^0 \pi^0$ (on the right) showing data (points), the combinatoric background (dashed red) and the full fit function  (solid blue) including signal and combinatoric background.}
\label{fig: DzToGG}
\end{center}
\end{figure}

\Dz candidates  are reconstructed in the decays $D^0 \ra \gamma \gamma$,   $D^0 \ra \pi^0 \pi^0$, and  $D^0 \ra \K0S  \pi^0$. In the decay   $D^0 \ra \gamma \gamma$ the photon candidates have a center-of-mass (\CM)  energy between 0.74 and 4 \gev and the main background $D^0 \ra \pi^0 \pi^0$ is suppressed with a $\pi^0$ veto. $B$ background in $D^0 \ra \gamma \gamma$ ( $D^0 \ra \pi^0 \pi^0$) is rejected selecting $D^*$ candidate with a \CM\  momentum above  $2.85$  ($2.4$)\gevc. 
Signal (reference mode) selection efficiency  for the  $D^0 \ra \gamma \gamma$ mode is $6.1$ ($7.6$) \%.  For the  $D^0  \ra \pi^0 \pi^0$  mode the signal (reference mode) selection efficiency is   $15.2$ ($12.0$) \%.  Signal yields are extracted  with unbinned Maximum Likelihood (\ML\ )  fit to the \Dz  invariant mass distribution (Fig.~\ref{fig: DzToGG}).

For $D^0 \ra \gamma \gamma$ the fit signal yield is $-6 \pm 15$.
Including statistical and systematic uncertainties the \UL at $ 90\%$  \CL\ for this decay mode is $2.4 \times 10^{-6}$. This preliminary  result is in agreement with \SM expectation. Based on this result enhancement due to gluino exchange cannot exceed a factor 70. 

 The preliminary  result for the branching fraction  of the $D^0 \ra \pi^0 \pi^0$ decay is $ (8.4 \pm 0.1 \pm 0.4 \pm 0.3) \times 10^{-4}$, where the uncertainties  refer  to statistical, systematic, and reference mode branching fractions uncertainties, respectively.  
 
\subsection{Search for $X_c \ra h l^+ l^-$}
In a recent  paper \cite{VeriRari} \babar\ searched for charm hadron decays of the type $X^+_c \ra h^{\pm} l^{\mp} l^{(')+}$ \cite{CC}, where $X^+_c$ is a charm hadron ($D^+, D^+_s$  or $ \Lambda^+_c$), $l^{(')+}$
is an electron or a muon, and $h^{\pm}$ can be a pion or a kaon (a proton in $\Lambda^+_c$ decay modes).  The analysis is based on a data sample of $384\, fb^{-1}$ of  \epem\ annihilation data collected at or close to the  $\Upsilon(4S)$ resonance.  

In total  35 decay modes have been studied. Among them there are  decay modes with oppositely  charged leptons of the same flavor   which proceed through \FCNC.  These decays are very rare. There are decays  with oppositely charged leptons of different flavor. These correspond to lepton-flavor violation  decay modes which  in \SM\  are essentially forbidden  because they can proceed  only through lepton mixing. There are decays  with two leptons of the same charge. These are lepton-number violating processes which are forbidden in the \SM.   The most stringent existing {\UL}s for  the branching fractions  of  the decays $X^+_c \ra h^{\pm} l^{\mp} l^{(')+}$ are in the range $[1 - 700] \times 10^{-6}$~\cite{Rubin, Abazov,Link,Aitala,Kodama}.

\newpage

 \begin{figure}[htb]
\begin{center}
\includegraphics[scale=0.20] {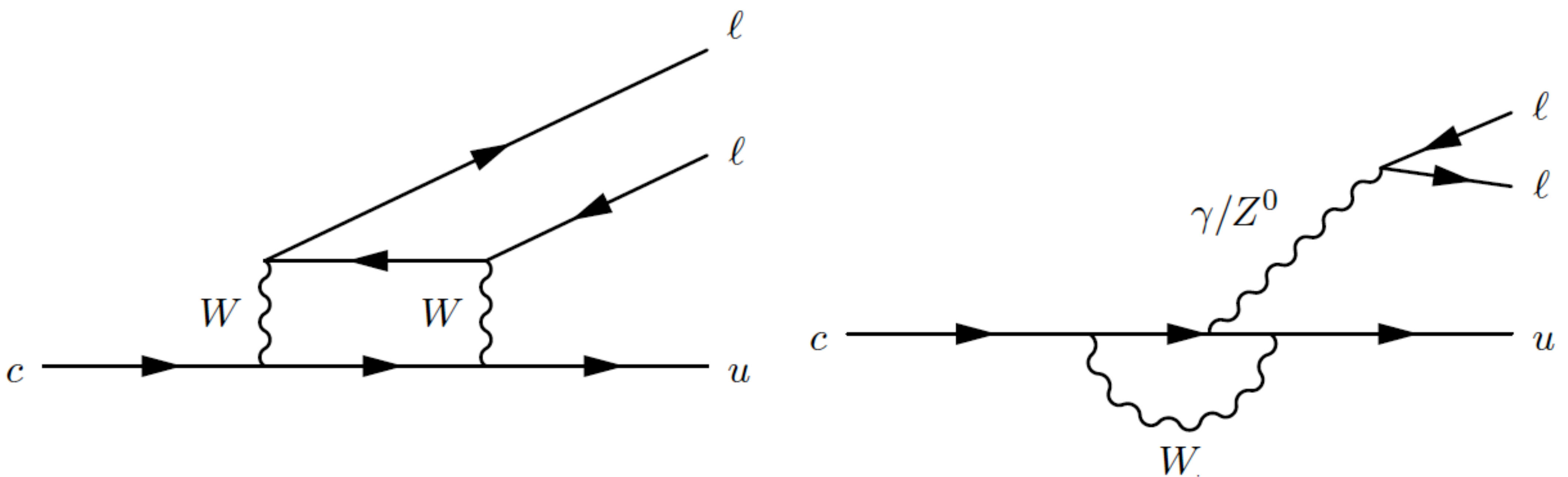} 
\vspace{-0.3cm}
\caption{ Standard Model short-distance contributions to the $c \ra ul^+ l^- transition$.}
\label{fig:DiagrRari}
\end{center}
\end{figure}

Transitions $c \ra u l^+ l^-$  proceed in \SM through diagrams shown in Fig.~\ref{fig:DiagrRari} and are expected with  branching fractions  of   ${\cal{O}}(10^{-8})$ \cite{Burdman, Fajfer2}.
Several extensions of  the \SM\  predict  an enhancement of  these branching fractions  \cite{Burdman, Fajfer2,Paul}.  The target  decay modes may have also long-distance contributions  from leptonic decays of intermediate resonances  like  $D_{(s)} \ra X_u V$ with $V \ra l^- l^+$ which are expected with a branching fraction  of   ${\cal{O}}(10^{-6})$ \cite{Burdman,Fajfer2}.  At current experimental sensitivity  these long-distance contributions can come only  from $D^+_{(s)}  \ra \pi^+ \phi $ decays with $\phi \ra l^- l^+$.  The effect of these  long-distance contributions is suppressed with   a cut  on the $l^- l^+$ invariant mass around the  $\phi$ meson.

In the signal  reconstruction  three tracks, one  identified as $\pi$, $K$, or proton  and two identified as electron or muon $(l l^{(')})$, are merged.  Charm hadron is selected with a momentum $p^*$ in  \epem  \CM\  $>  2.5$ \gevc  to suppress charm hadron production from $B$ decays. The final selection is done with a likelihood ratio (\LR) using three discriminating variables ($p^*$, total reconstructed energy in the event, and flight length significance of the charm hadron candidate).  The minimum \LR\  value is chosen independently for each decay mode. 

Signal yields are extracted with extended unbinned  \ML\  fit to the  invariant mass distributions of $hll$. There are  three components in the fit: signal, combinatoric background, and  background  from nonleptonic charm decays in which  two hadrons are misidentified as leptons.  The fitted signal yields are translated into branching fractions  by normalizing  them to the yields of known  charm decays with similar kinematics  ($D^+_{(s)}  \ra  \pi^+ \phi$ ($\phi \ra KK$)  and $\Lambda^+_c \ra p K^- \pi^+$).   No  signals are found  and Bayesian {\UL}s at 90\% \CL\  are calculated.  Limits   in 32 of the 35 studied decay modes are improved upon  the previous ones, in most cases by more than order of magnitude.

\section {Measurement of $|V_{ub}|$ from Inclusive $\Bbar  \ra X_u l \Nubar$ Decays}
The magnitude of  the Cabibbo-Kobayashi-Maskawa  quark-mixing matrix element  $V_{ub}$~\cite{CKM} can be determined from inclusive  semileptonic $\Bbar$ decays  to charmless final states $X_u l \Nubar$, where $ l = e $ or $\mu$,  and $X_u$ is a hadronic system (without charm).   The real difficulty in this inclusive measurement comes from the 
overwhelming  charm background from $ \Bbar \ra X_c l \Nubar$ which has a rate fifty times larger and an event  topology very similar to signal.  

In a recent analysis~\cite{VUB}  \babar, using the full dataset  of $467 \times 10^{6}$  \BBbar pairs,  has  measured  the partial branching fractions   of the
$\Bbar \ra X_u l \Nubar$ decays \cite{CC}, restricting  the analysis in selected regions  of the phase space where most  effective is the suppression of the charm background.  
The event selection uses a hadronic tag: in the sample of $\Upsilon(4S) \ra B \bar{B}$ events,  one $B$  decaying into hadrons is 
fully reconstructed ($B_{reco}$) and  the other $B$  ($B_{recoil}$) is identified by the presence of an electron or a muon.
The $B_{reco}$ is reconstructed in many exclusive hadronic decays  $B_{reco} \rightarrow  {D}^{(*)} Y^{\pm}$, where  $D^{(*)}$ is a charmed meson (\Dz, \Dp, \Dstarz, 
 or   \Dstarpm) while the  charged hadronic system $Y^{\pm}$ consists of up to five charged hadrons, pions or kaons plus up to two neutral mesons ($K^0_S$, $\pi^0$).

In the $B_{recoil}$ rest-frame we require only one charged  lepton with momentum $p^*_l > 1 $ \gevc\  and the hadronic system  $X_u$ is reconstructed from charged particles and neutral clusters not associated to the $B_{reco}$ or to  the charged lepton. Neutrino is reconstructed from the missing four-momentum in the whole event.  Requirements on several kinematic observables  were applied in different phase space regions to select the final signal events.  

Partial branching fractions are  measured in several regions of phase space and   are normalized to the total semileptonic branching fractions,  thus reducing  several  systematic uncertainties. Considering the most inclusive measurement   (based only on the requirement 
$p^*  >1.0$  \gevc),  from a two-dimensional  $M_X - q^2$ \ML\ fit to the hadronic invariant mass $M_X$ and the leptonic mass squared $q^2$, we measure:

\begin{equation}
\Delta {\cal B}(\Bbar  \ra X_u l \Nubar;  p^*_l > 1.0 \gevc) =  (1.80 \pm 0.13 \pm 0.15 \pm 0.02) \times 10^{-3} \,,
\label{equ=MI}
\end{equation}

\noindent where the first uncertainty is statistical, the second  systematic, and the third theoretical.

The measured partial branching fractions are related to $|V_{ub}|$ via the following relation:
$$
|V_{ub}| = \sqrt{\frac{\Delta{\cal B}(\Bbar \ra X_u l \Nubar)}{\tau_B \Delta \Gamma_{theory}}}\, ,
$$

where $\tau_B$ is the $B$ meson lifetime and  $\Delta \Gamma_{theory}$ is the theoretically predicted $\Delta{\cal B}(\Bbar \ra X_u l \Nubar)$  partial branching fraction  for the selected phase space region. This prediction is calculated  on the basis of  four  different QCD models: Bosch, Lange, Neubert, and Paz (BLNP) \cite{Neubert},  Gambino, Giordano, Ossola, and Uraltsev (GGOU) \cite{Gambino}, Andersen and Gardi (DGE) \cite{Gardi}, and Aglietti, Di Lodovico, Ferrera, and Ricciardi (ADFR) \cite{Aglietti}.  Results for $|V_{ub}|$ for these four different QCD calculations using the partial branching fraction obtained  in the most inclusive measurement  
(Eq.~\ref{equ=MI}) are presented in Table~\ref{tab:Vub}.  In the last row of Table~\ref{tab:Vub} we give the arithmetic average of the values and uncertainties obtained with the four QCD calculations (the first uncertainty is experimental and the second theoretical).  The total uncertainty is about $6.9$ \%, comparable in precision with the Belle result~\cite{InclBelle}.

\begin{table}[htb] 
\caption{ Results for $|V_{ub}|$  for the four different QCD calculations for the most inclusive partial branching fraction measurement.}
\label{tab:Vub}
\begin{center}
\begin{tabular}{lc}
\hline\hline
QCD Calculation                   & $|V_{ub}| (10^{-3})$       \\\hline
BLNP                                       & $4.27 \pm 0.15 \pm 0.18^{+0.23}_{-0.20}$                                       \\ 
DGE                                         & $4.34 \pm 0.16 \pm 0.18^{+0.22}_{-0.15}$                                      \\
GGOU                                      & $4.29 \pm 0.15 \pm 0.18^{+0.11}_{-0.14}$                                       \\
ADFR                                       & $4.35 \pm 0.19 \pm 0.20^{+0.15}_{-0.15}$                                       \\\hline\hline
 Arithmetic Average                &  $4.31 \pm 0.25 \pm 0.16$          \\
\hline\hline
\end{tabular}
\end{center}
\end{table}

The value   of  $|V_{ub}|$ obtained  in this inclusive analysis is higher than the value obtained by \babar\ in an exclusive analysis \cite{VUBE}. The discrepancy is at a level  of about $2.7 \, \sigma$ and is 
increased in the latest measurements obtained with significantly decreased uncertainties thanks to the improved experimental techniques and theoretical inputs.  A similar discrepancy  is also  present  in BELLE results \cite{InclBelle,ExclBelle}.

\section {Measurement of the decays $\Bbar \ra D  \tau^- \Nubar_{\tau}$ and $\Bbar \ra D^{*} \tau^- \Nubar_{\tau}$} 
Semileptonic B decays  to $\tau$  lepton $\Bbar \ra D^{(*)} \tau^- \Nubar_{\tau}$ 
 \cite{CC} can provide constraints on the \SM\ \cite{Korner,Adam,Hwang} and are sensitive to physics beyond the \SM.  In extensions of 
the \SM (such  as in the multi-Higgs doublet models  and the \MSSM ), intermediate charged Higgs boson can contribute to the amplitude, modifying significantly 
the branching fraction ~\cite{Bohdan,Tanaka,Ken,Itoh,Chen,Nierste1}.  Branching fractions of these  decay modes are smaller compared to those of decay modes to final states  containing a light lepton, l = e or $\mu$. 
\ SM\  expectations for the relative rates  between signal and  reference modes are:    \calR =  ${\cal B}(\AntiBDtauNu)/{\cal B}(\AntiBDlNu) = 0.31 \pm 0.02$  \cite{Nierste2} and    \calRst  =  $ {\cal B}(\AntiBDsttauNu)/{\cal B}(\AntiBDstlNu) =  0.25 \pm 0.02$ \cite{Chen}. Multi-Higgs doublet models predict an effect  on the ratio \calR\  much stronger than on \calRst ~\cite{Bohdan,Tanaka,Ken,Itoh,Chen,Nierste1}. This effect may enhance or decrease ${\cal R}(D^{(*)})$,  depending  on the value of  the ratio $\tan\beta$/$m_{H^{\pm}}$ of the Higgs parameter $\tan \beta$ and the charged Higgs mass $m_{H^{\pm}}$.

Recently ~\cite{Sevilla} \babar\  updated with the full data sample  of $471 \times 10^{6} $  \BBbar pairs previous analyses  \cite{PrevBaBar}  of the decays $\Bbar  \ra  D^{(*)} \tau \Nubar_{\tau}$.  $\UfourS \ra \BBbar $ events are tagged by the hadronic decay  of one of the $B$ mesons  ($B_{tag})$.  The tag decay modes are of the type 
$B_{tag} \ra S X^{\pm}$, where S can be  $D$, $\Dst$, $D_s$, $D^*_s$, or $J/\psi$  reconstructed in many s. $X^{\pm}$ is a charged state with a maximum of 5 particles ($\pi$ or K) including up to two neutral particles ($\pi^0$ or $K^0_S$).  

For each $B_{tag}$ candidate in a selected event,  the other $B$($B_{sig})$  is searched for combining a single charged lepton and a $D^{(*)}$ meson.
The $\tau$ lepton is reconstructed only in the purely leptonic decays $\tau^-  \ra e^- \Nubar_e \nu_{\tau}$ and  $\tau^-  \ra \mu^- \Nubar_{\mu} \nu_{\tau}$ while the $D^{(*)}$ meson  is reconstructed 
in the four modes  $D^0$, $D^{*0}$, $D^+$, and $D^{*+}$.   \Dz (\Dp)  is reconstructed in 5 (6)  decay modes for a combined branching fraction of $35.8$\% ($27.3$\%). \Dst meson is identified in the decays $D^{*+} \ra \Dz \pi^+$, $\Dp \pi^0$ and  $D^{*0} \ra \Dz \pi^0$, $\Dz \gamma$.
The signal modes have in the final state one secondary lepton and three neutrinos while the reference modes have a primary lepton and  one neutrino.

For the separation of signal and reference modes  the most discriminating variable is the missing mass squared , defined as $m^2_{miss} = (p_{e^+ e^-} - p_{tag} -p_{D^{(*)}} - p_l )^2  $, where $p$ are four-momenta. This quantity  peaks at zero in the reference decay modes where only one neutrino is missed while in the signal modes the  $m^2_{miss}$ distribution is broad and extends up to 8 $(\gevcc)^2$. Further separation of signal and reference modes is obtained with a requirement on  the minimum momentum transfer . For decays with a $\tau$, $q^2 = (p_{\tau} + p_{\nu_{\tau}})^2 > m^2_{\tau} \cong 3.16$ $(\gevcc)^2$.  The applied requirement $q^2  > 4 $ $(\gevcc)^2$  retains $98$\% of the signal decays and rejects more than $30$\% of the reference modes. 

\begin{table}[htb]
\caption{Results for the ratios \RDx,  the individual signal branching fraction,  and the signal significance $S_{tot}$ including systematic uncertainties, and the significance $S_{stat}$ (only statistical uncertainties) }
\label{tab:FinalResults}
\begin{center}
\begin{tabular}{lccc}
\hline\hline
Decay Mode                           & \RDx                                              &${\cal B}(B\to\ds\tau\nu)\,(\%)$        &  $S_{tot}$ ($S_{stat}$) \\ \hline
$\Dz\taum\nutb$                     & $0.422  \pm 0.074  \pm 0.059$      & $0.96 \pm 0.17 \pm 0.14$        & $5.0\,  (6.2) $                       \\ 
$\Dstarz\taum\nutb$               & $ 0.314 \pm 0.030 \pm  0.028$      &$ 1.73  \pm 0.17 \pm 0.18$       & $8.9\,  (11.9) $                        \\
$\Dp\taum\nutb$                     & $0.513 \pm 0.081 \pm 0.067$         & $1.08  \pm 0.19  \pm 0.15$        & $6.0\,  (7.5) $                        \\
$\Dstarp\taum\nutb$              & $0.356 \pm 0.038 \pm 0.032$         &$1.82  \pm 0.19  \pm 0.17$        & $9.5\,  (12.1) $                        \\ \hline
$D\taum\nutb$                        & $0.456 \pm 0.053 \pm 0.056$         &$1.04  \pm 0.12  \pm 0.14$        & $6.9\,  (9.6) $                        \\
$\Dstar\taum\nutb$                 & $0.325 \pm 0.023 \pm 0.027$        &$1.79  \pm 0.13  \pm 0.17$         & $11.3\,  (17.1)$                        \\
\hline\hline
\end{tabular}
\end{center}
\end{table}

Combinatorial background is reduced constraining  to the same vertex the charged daughters of the $D^{(*)}$ and $B$ mesons.   Improved discrimination of signal and reference modes from  several backgrounds is obtained using   a multivariate discrimination with  a Boosted Decision Tree (\BDT): 12 discriminating classifiers using 8 discriminating variables each.

Source of difficult background is the decay  $\Bbar \ra D^{**} l^- \Nubar_l$.  $D^{**}$ mainly decays to $D^{(*)} \pi$ which enter in the selection when the pion is neutral and not reconstructed or is charged  and  associated to $B_{tag}$. This background has been studied with four control samples  ($D^{(*)} \pi^0$) identical to the signal, except for an additional $\pi^0$ selected in the mass range $[120\,,150]$ \mev.

 To extract yields for signal and reference modes we perform an extended, unbinned \ML\   2D  fits to the distribution of $m^2_{miss}$  vs $|p^*_l|$  (lepton momentum calculated in the B meson rest frame). Simultaneous fit to the four signal channels (\Dz, $D^{*0}$, \Dp, $D^{*+}$) and the four  $D^{(*)} \pi^0$ channels.  Another fit is performed  imposing the isospin constraint: ${\cal R}(D^0) =  {\cal R}(D^+) \equiv   {\cal R}(D)$   and ${\cal R}(D^{*0}) =   {\cal R}(D^{*+}) \equiv   {\cal R}(D^*)$.  
 
The fit describes reasonable well the four  $D^{(*)} \pi^0$ channels inside the sizable statistical uncertainties. The fit describes very well the large contributions of the reference decay modes.
Fit results for $D^{*0}$ and $D^{*+}$  channels are very good. Both channels are observed with a significance higher than $11\, \sigma$ (only statistical uncertainties).  For the $D$ channels the fit 
projection onto  $m^2_{miss}$  shows an excess of data in the range $ 1.5 < m^2_{miss} < 3.5$ $(\gevcc)^2$ and an overestimate of events for $m^2_{miss} > 5$ $(\gevcc)^2$. These regions are dominated by continuum and $B$ combinatorial backgrounds which are fixed in the fit to what is expected by simulation. It  is not clear if the fit-data  differences    are  statistical or systematic. Both \Dz\  and \Dp\ channels are observed with a significance (statistical uncertainties only) higher than $6 \, \sigma$.
 
 Main sources of systematic uncertainties are the MC simulation of the background, the statistical uncertainties of the simulated samples, and the $D^{**} l \nu$ decay modes.  In order to understand the origin of the difference in the fit-data  comparison, the \BDT\  requirements have been changed in such a way to have a sample passing the selection $50$\%. $80$\%, $120$\%, and $200$\% compared to the nominal sample.  The agreement between fit and data improved using both more and less restrictive \BDT\  requirements.  We assign a  systematic uncertainty equal to half of the variation on ${\cal R}(D^{(*)}$ in the fit when applying tight BDT requirements ($50$\% nominal sample) and loose \BDT\  requirements ($200$\% nominal sample). This systematic uncertainty  ($9.5$\% on ${\cal R}(D)$ and $6.5$\% on   ${\cal R}(D^*)$) is   comparable in size with statistical uncertainties. It should be eliminate or reduced once the source of the difference in fit-data is understood. 

 Table \ref{tab:FinalResults} summarizes the results obtained from the two fits: the first in which all four signal yields can vary independently, and the second (last two rows in the table) in which isospin relations are imposed.  

These preliminary results are in agreement  with previous BaBar results \cite{PrevBaBar} and with  Belle measurements \cite{PrevBelle} and have significantly reduced uncertainties. The decays $\Bzb \ra D^{+} \tau^- \Nubar_{\tau}$ and $B^- \ra  D^0 \tau^- \Nubar_{\tau}$ are observed for the first time.

\newpage

\begin{figure}[h]
\begin{center}
\includegraphics[scale=0.20] {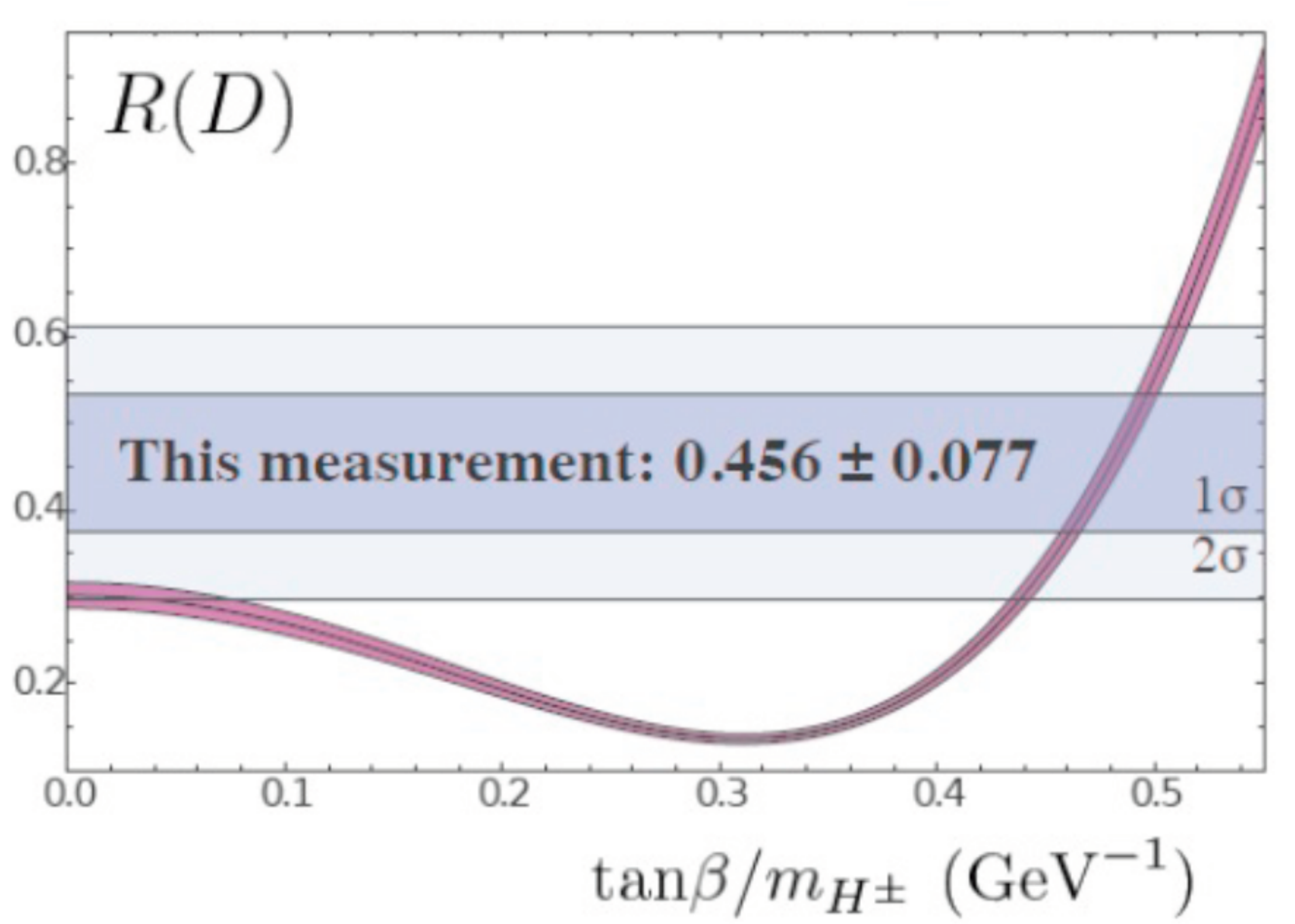} 
\caption{${\cal R}(D)$ as a function of the ratio  $\tan \beta$/$m_{H^{\pm}}$. }
\label{fig:Tanaka}
\end{center}
\end{figure}

Figure~\ref{fig:Tanaka}  shows  ${\cal R}(D)$ as a function of the ratio  $\tan \beta$/$m_{H^{\pm}}$. The violet band is the theoretical prediction \cite{Tanaka2} while the horizontal blue band is the present measurement.  Present result favors large values of  $\tan \beta$/$m_{H^{\pm}}$.  Furthermore ${\cal R}(D)$ is about $1.8\, \sigma$ in excess over \SM\  prediction. A similar excess is also measured for ${\cal R}(D^*)$. 

 \section{Direct CP Asymmetry in Inclusive $B \ra X_{s+d} \gamma$}
\begin{figure}[h]
\begin{center}
\includegraphics[scale=0.40] {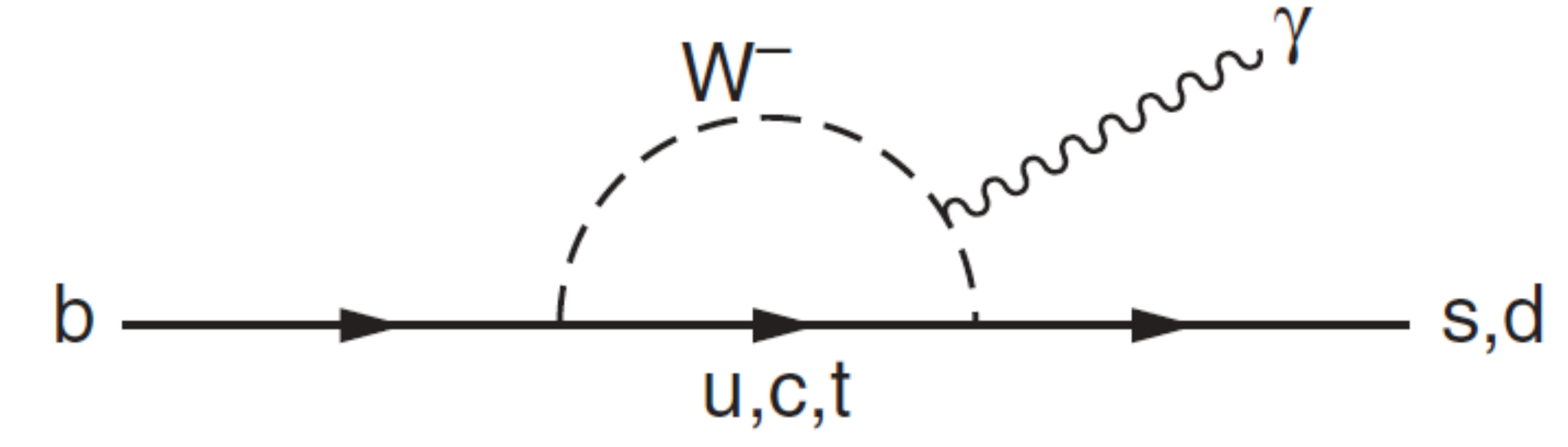} 
\caption{ Leading order diagram for $b \ra (s,d) \gamma$.}
\label{fig:Radiative}
\end{center}
\end{figure}

In the \SM\  the inclusive electromagnetic radiative decays  $b \ra s\gamma$ or $b \ra d\gamma $ proceed at  the leading order Feynman diagram  shown in Fig.~\ref{fig:Radiative}.

 The branching fraction of the process  ${\cal B}(B \ra X_s \gamma) =  (3.15 \pm 0.23) \times 10^{-4}$ with $E_{\gamma} > 1.6$  \gev (in the $B$ meson rest frame) has been calculate  in \SM at next-to-next-leading order with a precision of $7$\%  \cite{Misaim}. Because  new heavy particles may enter in the loop at leading order,  the value of this branching fraction is highly sensitive to \NP\  \cite{Haisch}.
Another  quantity highly sensitive to contributions of \NP\ is the direct \CP asymmetry \cite{Kagan} defined as: 
 
 $$
 {\cal A}_{\CP}  = \frac{\Gamma(b \ra s\gamma + b \ra d\gamma) - \Gamma(\bbar \ra \sbar \gamma + \bbar \ra \dbar\gamma)} {\Gamma(b \ra s\gamma + b \ra d\gamma) +\Gamma(\bbar \ra \sbar \gamma + \bbar \ra \dbar \gamma)}
 $$
 
 This asymmetry in \SM is expected $\approx 10^{-6}$  \cite{Kagan} with the asymmetries for $b\ra s\gamma$ and $b \ra d\gamma$  opposite  with nearly exact cancellation.
 In \NP scenarios  ${\cal A}_{\CP}$  can be at  about  $10$\% \cite{Hurth}. 
 
 \newpage
 
\begin{figure}[ht.]
\begin{center}
\includegraphics[scale=0.40] {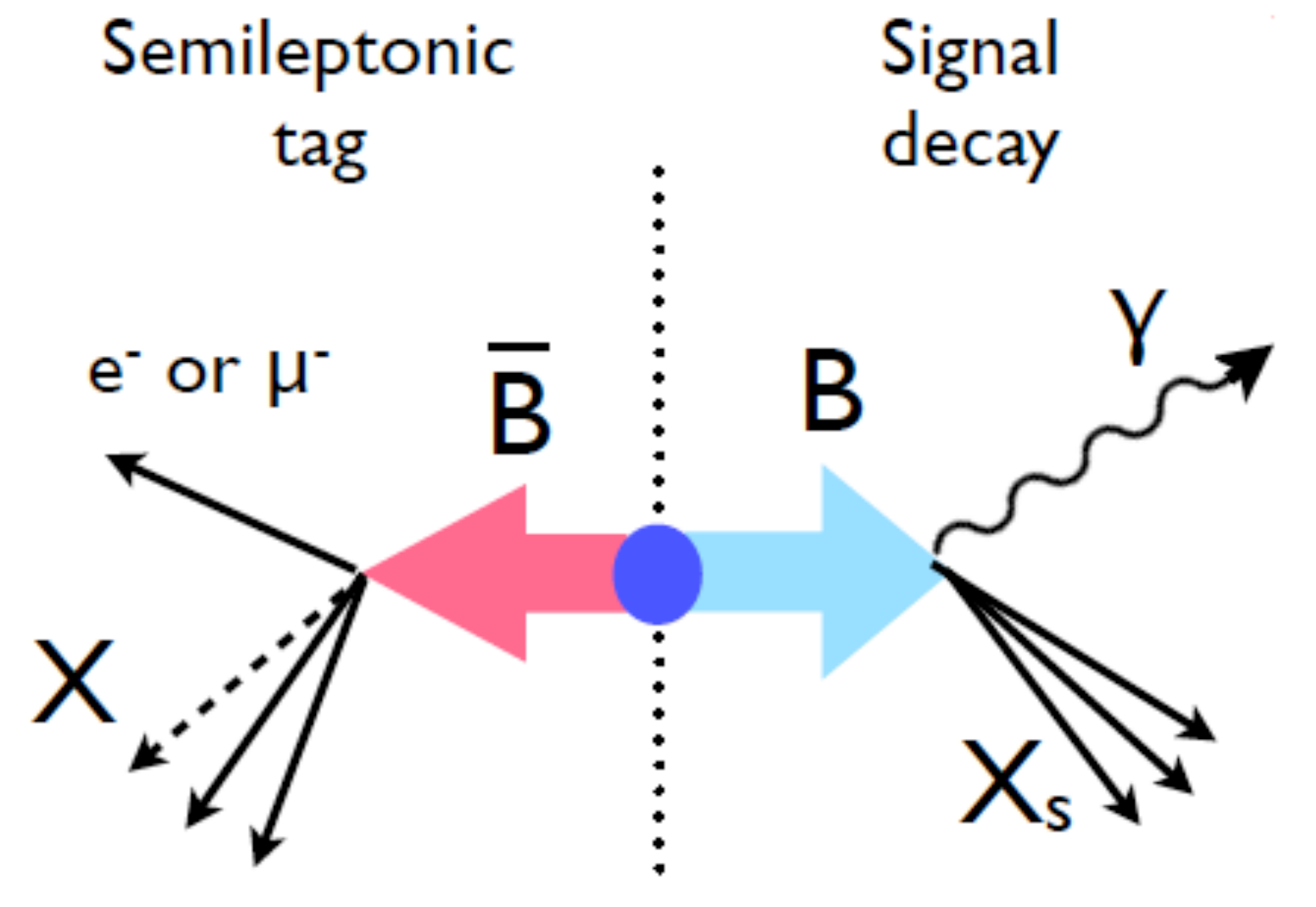} 
\caption{Semileptonic tag. }
\label{fig:SemTag}
\end{center}
\end{figure}

 In a recent  analysis  \babar\  has measured direct \CP\ asymmetry  in  inclusive $B \ra X_{s+d} \gamma$ using a data sample of $383 \times 10^6$ \BBbar\   pairs. The reconstruction of the event uses a fully inclusive method with a semileptonic tag (Fig.~\ref{fig:SemTag}).

  Signal events are identified by high energy photon and the $X_s$ is not reconstructed.  Therefore is not possible to distinguish $X_s$ and $X_d$ states and  what is measured is  $B \ra X_{(s+d)} \gamma$.    The $B_{tag}$ is searched for in a semileptonic  decay mode, checking  that the remaining particles in the event are consistent with a B decay. Photon selection requires at least one photon with energy $ 1.53 < E^*_{\gamma}  < 3.5 $ \gev\ and the tag lepton can be an electron or a muon with momentum  $p^* > 1.05$ \gev ($E^*_{\gamma}$ and $p^*$ are calculated in the \UfourS\ rest frame).  
 Continuum background is suppressed using a neural network discriminant based on eight topological variables. Remaining  continuum background  in the final sample is estimated   using off-resonance data collected $40$ \mev\ below the \UfourS\ resonance. The \BBbar background  is mostly due to photons from low-mass mesons (mainly $\pi^0 $ and $\eta$). This background is  removed  using explicit vetoes. Remaining \BBbar background is estimated from MC simulation and  is cross-checked against data with control samples.
Continuum and \BBbar background estimation have been validated with control samples provided by the photon spectrum. 
 
\begin{figure}[h]
\begin{center}
\includegraphics[scale=0.33] {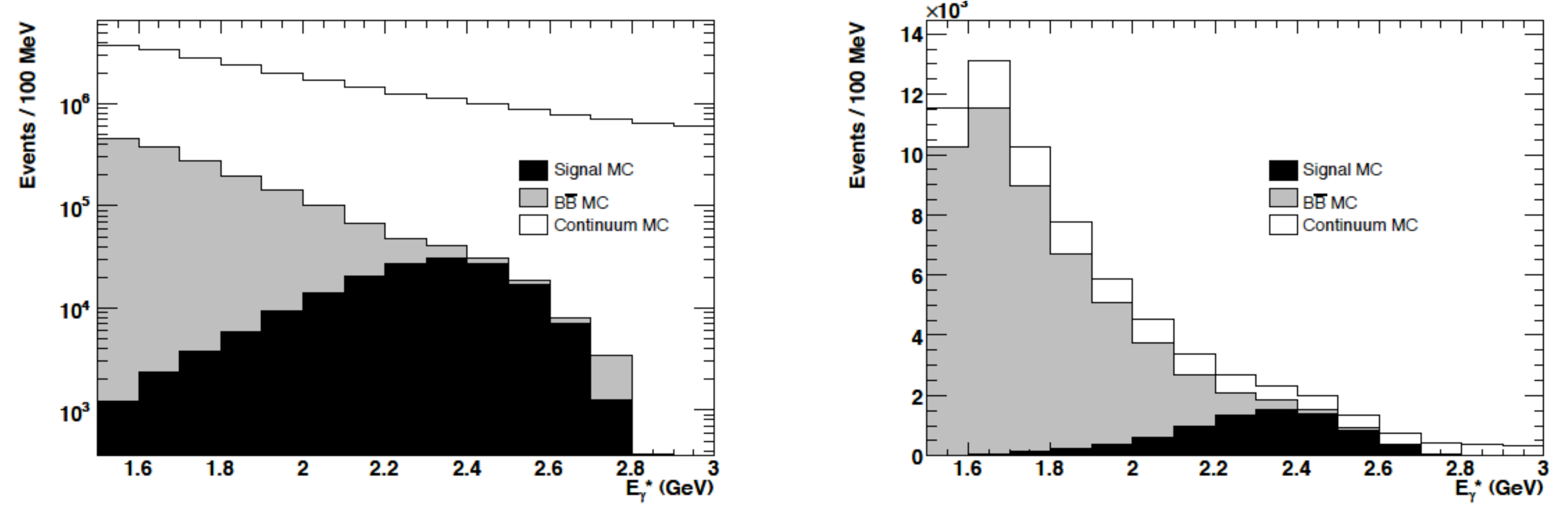} 
\space{-0.4cm}
\caption{Photon spectrum after requiring a high energy photon on the left and after all selection requirements on the right. }
\label{fig:Selection}
\end{center}
\end{figure}

 In Fig.~\ref{fig:Selection} we show the signal and  backgrounds distributions with only the high energy photon  requirement (on the left in logarithmic  scale) and  after all selection requirements (on the right in linear scale). 
  
\newpage
 Direct \CP asymmetry is insensitive to the photon energy cut. So to reduce the sensitivity to background the asymmetry is calculated in the optimized range ($2.1 - 2.8$) \gev. Lepton charge gives B flavor and separation of  $B$ and  $\Bbar$   is done  according to the charge of the lepton tag.  The tagged signal yields are $N^+ = 2623 \pm 158$ and $N^- = 2397 \pm 151$  and the measured \CP\ asymmetry is: 
  
 $$
 {\cal A}^{meas}_{\CP}  =  \frac{N^+ - N^-}{N^+ + N^-} = 0.045 \pm 0.044
 $$ 
 
This result must be corrected for the dilution due to  mistag fraction $\omega$, $ {\cal A}_{\CP} = \frac{{\cal A}^{meas}_{\CP}}{1 - 2 \omega}$, 
 for the uncertainty in the \BBbar background estimation, and for the bias induced by \CP asymmetry  in the  \BBbar  background and by   detection asymmetry.  Correcting $ {\cal A}^{meas}_{\CP} $ for these effects we obtain the preliminary result:
 
 $$
 {\cal A}_{\CP}  = 0.056  \pm 0.060_{stat} \pm 0.018_{syst}
 $$
 
 No significant asymmetry is observed in agreement with \SM expectations. This result is the most precise to date \cite{Rad}.

\section {Search for \CP Violation in $D^+ \ra \K0S \pi^+$ }
 \CP   violating asymmetries have been measure both in $K$ and $B$ systems with results in agreement with SM expectations. \CP violation has yet not been observed in  charm decays,  where 
 the \SM\  expectations for CV violating asymmetries  are at the level of $10^{-3}$  or less \cite{Buccella}.   

 In a recent analysis \cite{K0Pi}  \babar\  searched for \CP violation in the decay $D^{\pm} \ra \K0S \pi^{\pm}$, measuring the direct \CP violating parameter:

$$
\calA_{\CP} = \frac{\Gamma(D^+ \ra \K0S \pi^+) - \Gamma(D^- \ra \K0S \pi^-)}{\Gamma(D^+ \ra \K0S \pi^+) + \Gamma(D^- \ra \K0S \pi^-)}
$$

with $\Gamma$ partial decay width of the decay.

Although the \SM prediction for direct \CP violation due to the interference between Cabibbo-allowed and doubly Cabibbo-suppressed amplitudes 
is negligible \cite{Lipkin},  \Kz - \Kzb mixing induces a time-integrated \CP violating asymmetry of $(-0.332 \pm 0.006) \%$ \cite{Nakamura}.
Contributions of physics beyond the \SM  may enhance the value of  $\calA_{\CP}$  up to 1\%  \cite{Lipkin, Bigi}. Previous measurements of $\calA_{\CP}$ have been reported  by CLEO-c [$-0.6 \pm 1.0 \pm 0.3$ \%] \cite{CLEOc}  and Belle Collaboration [$-0.71 \pm 0.19 \pm 0.20$\%] \cite{BelleK0SPi}

We reconstruct  $D^{\pm}_s  \ra \K0S \pi^{\pm}$ decays combining a \K0S  candidate with a charged pion candidate.  The \K0S decays to $\pi^+ \pi^-$ and is selected with an invariant mass within $\pm 10$ \mevcc of the nominal \K0S mass \cite{Nakamura}. The reconstructed pion candidate is selected with a momentum $p_T$ in the plane perpendicular to the z axis greater than 400 \gevc. The selected $D^{\pm}$ candidate has an invariant mass within $\pm 65$
\mevcc of the nominal $D^+$ mass \cite{Nakamura}, and a momentum   $p^*$  in the   \epem\  CM  in the range  ($2 - 5$) \gevc. Further suppression of background  has been achieved using a \BDT\  with seven event topological discriminating variables.

Signal yield is extracted with a binned  \ML\  fit to the invariant mass distribution of the selected $D^{\pm}$ candidates. There are three fit components: signal, a background from $D^{\pm}_s  \ra \K0S K^{\pm}$ and a combinatoric background.

\newpage

 \begin{figure}[htb]
\begin{center}
\includegraphics[scale=0.30] {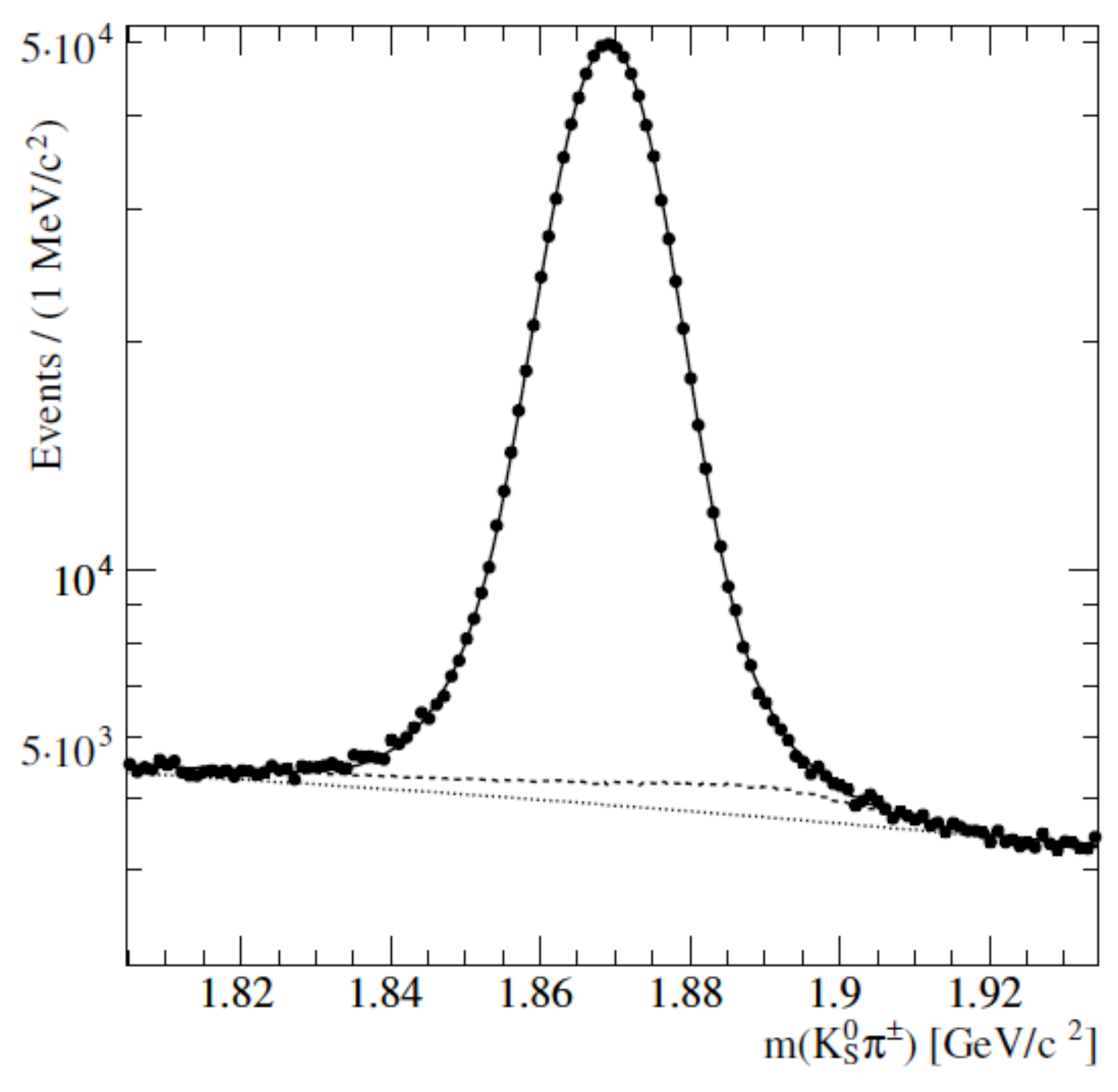} 
\caption{Invariant mass distribution for $\K0S \pi^{\pm}$: solid curve is the fit to the data (points), the dashed line is the sum of backgrounds, and the dotted line is the combinatorial background only. The vertical scale on the plot is logarithmic. }
\label{fig:InvMass}
\end{center}
\end{figure}

Data and fit are shown in Fig.~\ref{fig:InvMass}. The signal yield asymmetry is:

$$
\calA= \frac{N_{\Dp} -N_{\Dm}}{N_{\Dp} + N_{\Dm}}\, ,
$$ 

where $N_{\Dp}$ and  $N_{\Dm}$ are the fitted yields for  $\Dp \ra \K0S \pi^+$ and $\Dm \ra \K0S \pi^-$, respectively.

The  quantity  $ \calA$  includes contribution not only from $\calA_{\CP}$ but also from the forward-backward  asymmetry $\calA_{FB}$ in $\epem\  \ra c \cbar$  due to 
$\gamma^* - Z^0$ interference and other QED processes.  Another source of asymmetry  ($\calA_{\epsilon}$)  contribution to \calA\ is induced by the detector as a consequence of  the difference in     reconstruction efficiency of    $\Dp \ra \K0S \pi^+$ and $\Dm \ra \K0S \pi^-$  due to differences in reconstruction and identification efficiencies  for $\pi^+$ and $\pi^-$. So we can write:

\begin{equation}
\calA = \calA_{\CP} + \calA_{FB} + \calA_{\epsilon}
\label{equ=AcpAfb}
\end {equation}

$\calA_{\epsilon}$ has been measured using a control sample of \BBbar\ decays. The bias of +$0.05$\% to  $\calA_{\CP}$ has been included in the systematics. 

We  separate    $\calA_{\CP}$  and $\calA_{FB}$ in Eq.~\ref{equ=AcpAfb}  considering that  $\calA_{FB}$  is  an odd function of $\cos\theta^*_D$, where $\theta^*_D$ is the polar angle  of the $D^{\pm}$ candidate momentum  in the \epem\   CM, while  $\calA_{\CP}$ is an even function of $\cos\theta^*_D$. Therefore the two asymmetries $\calA_{\CP}$ and $\calA_{FB}$ can be written as a function of $|\cos\theta^*_D|$ as:

$$
\calA_{FB}(|\cos\theta^*_D|)  = \frac{A(+|\cos\theta^*_D|) - A(-|\cos\theta^*_D|)}{2}  
$$

$$
\calA_{\CP}(|\cos\theta^*_D|)  = \frac{A(+|\cos\theta^*_D|) + A(-|\cos\theta^*_D|)}{2}  
$$

The selected sample is divided in subsamples corresponding to 5 bins of  $|\cos\theta^*_D|$
  and a simultaneous binned \ML\  fit is performed to extract signal yield asymmetries. 
  
  \newpage
  
\begin{figure}[htb]
\begin{center}
\includegraphics[scale=0.45] {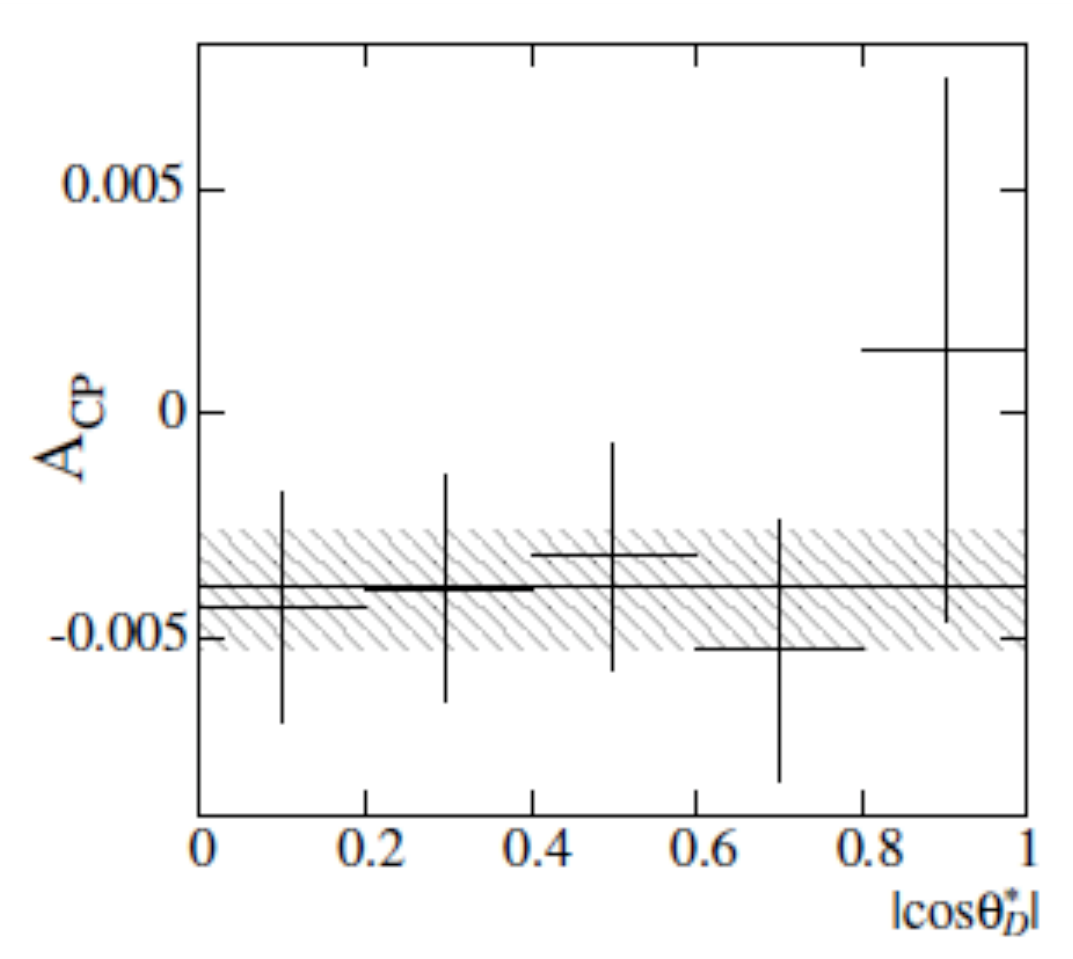} 
\hspace{1.5cm}
\includegraphics[scale=0.45] {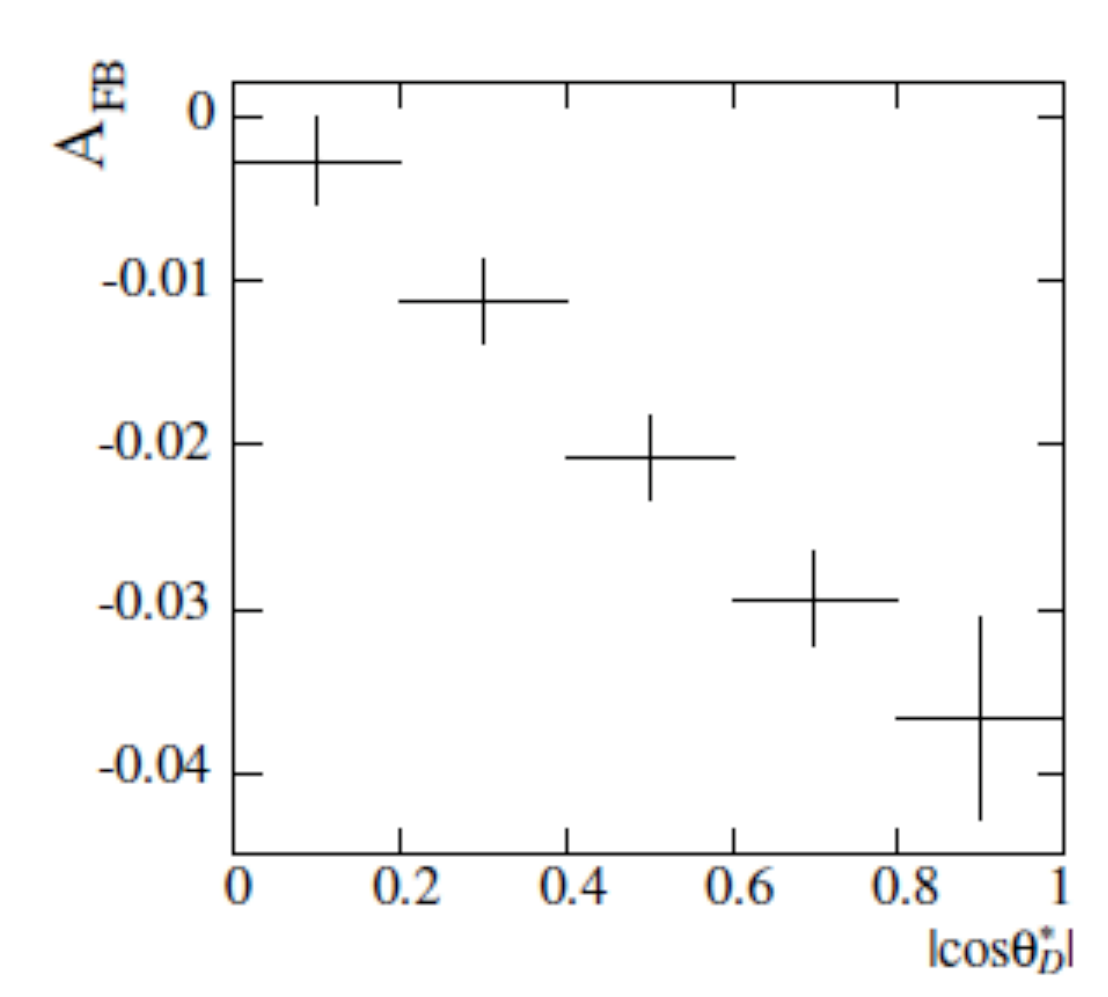} 
\vspace{0.5cm}
\caption{On the left $\calA_{\CP}$  with the central value (solid line)  and   the  $\pm 1 \sigma$ interval (hatched region), and  $\calA_{FB}$ (on the  right) in bins of 
$|\cos\theta^*_D|$. }
\label{fig:fit}
\end{center}
\end{figure}

  The five measured values of the parameters   $\calA_{\CP}$  and  $\calA_{FB}$ are shown in   Fig.~\ref{fig:fit}.   We measure:
   
  $$
  \calA_{\CP} = (-0.39 \pm 0.13 \pm 0.10)\% \, ,
  $$
  where the first  uncertainty is statistical  and the second systematic. This value is in agreement with the prediction  of the \SM.

\section {Search for T Violation using T-odd Correlations in $D^+_{(s)}  \ra K^+ \K0S  \pi^+ \pi^-$}
T violation in the channels $D^+ \ra K^+ \K0S \pi^+ \pi^-$  and $D^+_s \ra K^+ \K0S  \pi^+ \pi^-$  \cite{CC} can be measured  using T-odd correlations.  Using the momenta of the final particles in the $D^+_{(s)}$ rest frame, the T-odd correlation observable can  be written as:

$$
C_T \equiv {\vec p}_{K^+} \cdot ({\vec p}_{\pi^+} \times  {\vec p}_{\pi^-})
$$ 

This triple product is odd under time reversal. Assuming CPT invariance, T violation is equivalent to \CP violation. We measure the asymmetry:

$$
A_T \equiv  \frac{\Gamma(C_T >0) - \Gamma(C_T <0)}{\Gamma(C_T> 0) + \Gamma(C_T<0)}\, ,
$$

 where $\Gamma $ is the decay rate of the process. 

Because Final State Interactions (FSI)  can  produce  an asymmetry $A_T $ different from  zero  \cite{Bigi2}, we measure also the T-odd asymmetry  for the \CP-conjugate decay process:

$$
\overline{A}_T \equiv  \frac{\Gamma(- \overline{C}_T >0) - \Gamma(- \overline{C}_T <0)}{\Gamma(- \overline{C}_T> 0) + \Gamma(\overline{C}_T<0)} \, ,
$$

where $\overline{C}_T \equiv  \vec{p}_{K^-} \cdot (\vec{p}_{\pi^-}  \times \vec{p}_{\pi^+})$ with momenta calculated  in the $D^-_{(s)}$ rest frame. To remove FSI  effects we measure  the quantity $\calA_T$:

 \begin{equation}
\calA_T  \equiv \frac{1}{2} (A_T -  \overline{A}_T)
 \label{equ:Asym}
 \end{equation}

which  is an asymmetry characterizing   T violation in the weak process  \cite{Bensalem}.

In a   previous \babar\ analysis \cite{Todd}  done on  the neutral decay  $D^0 \ra K^+ K^- \pi^+ \pi^-$   no evidence of \CP violation has been found.
 
We describe here the search for \CP\   violation  using T-odd correlations in the decays $D^+  \ra  K^+ \K0S  \pi^+ \pi^-$ and $D^+_{s}  \ra K^+ \K0S  \pi^+ \pi^+$  \cite{Maurizio}.  In the reconstruction we have considered inclusive $D$ decays ($e^+e^- \ra X  D^+_{(s)} $), selected using kinematic constraints and particle 
identification.  
In the vertex  fit, requiring that all three tracks  with net charge +1  originate from a common vertex, a $\chi^2$ fit  probability $P1 > 0.1$ \%  has been  imposed. A second fit constraining the three tracks to originate from the \epem\ interaction region was also done.  In this case the  $\chi^2$ fit probability $P2$ is large for most background events with  tracks originating  in the interaction region.

$D^+_{(s)}$  candidate must have a momentum $p^*$ in the CM greater than 2.5 \gevc . To improve signal and background separation we consider the signed transverse decay length:

$$
L_T = \frac{\vec{d} \cdot \vec{p}_T}{|\vec{p}_T|}\, ,
$$ 

where $\vec d$ is the distance vector between IR and the $D^+_{(s)}$ decay vertex in the transverse plane and ${\vec p}_T$  is the transverse momentum of $D^+_{(s)}$. With the variables $p^*$, P1-P2, and $L_T$ we construct a \LR\  to optimize the signal yields separately for $D^+$ and $D^+_s$. Separation requirements  where optimized maximizing  the statistical significance $S/\sqrt{S + B}$, where $S$ is the number of signal and $B$ the number of background events in the signal region.

Then the dataset is divided in four  samples  depending of the $D^+_{(s)}$ charge and sign of  $C_T$ ($\overline{C}_T$) and on these four datasets a simultaneous fit to mass spectra  has been done to extract  the asymmetry parameters
$A_T$ and $\overline{A}_T$. The measured values are:

\begin{eqnarray}
A_T (D^+) &=& (+11.2 \pm 14.1 \pm 5.7) \times 10^{-3}\nonumber \\ 
\overline{A}_T (D^-)& = &(+35.1 \pm 14.3 \pm 7.2) \times 10^{-3}  \nonumber \\
A_T (D^+_s) & = & (+99.2 \pm 10.7 \pm 8.3) \times 10^{-3} \nonumber \\
 \overline{A}_T (D^-_s) & = &(+72.1 \pm 10.9 \pm 10.7) \times 10^{-3} \nonumber \,  ,
\end{eqnarray} 

where the first uncertainty is statistical and the second systematic.
FSI Effects are larger in $D^+_s$ than in $D^+$ decays. A study of such effects can be found in ref.~\cite{Gronau}. The T violation asymmetries  obtained using Eq.~\ref{equ:Asym} are:

\begin{eqnarray}
\calA_T(D^+) &=& (-12.0 \pm 10.0 \pm 4.6) \times 10^{-3} \nonumber \, \\
\calA_T(D^+_s) &= &(-13.6 \pm 7.7 \pm 3.4) \times 10^{-3} \nonumber 
\end{eqnarray} 

T violation parameter is consistent with 0 for both the two decay modes.

\section{Summary and Conclusions} 
We have presented \babar\ results of  recent    searches  for rare or forbidden  charm decays , measurement of the magnitude of the CKM matrix element  $V_{ub}$,  the  observation of the    semileptonic $B \ra D^{(*)} \tau \nu_{\tau}$ decays,  searches for  direct \CP\ violation asymmetry in $B\ra X_{(s+d)}\gamma$ and in $D^+\ra K^0_S \pi^+$, and T-violation in $D^+_{(s)} \ra K^+ \K0S  \pi^+ \pi^-$.  All these analyses use the final \babar\  dataset.   Results have been improved over previous analyses and are   
all  consistent with \SM\ expectations.  No \CP\ violation has been observed in $c \ra s$ transitions , neither from \SM\  nor from \NP.  More stringent limits are set on space parameters of \NP\ models.

B-factories reached their sensitivity limit. We expect soon a significant impact in flavor physics from LHCb experiment  \cite{LHCb} and (in a few years) from the next generation super B factories (SuperB \cite{superB} at the Cabibbo Laboratory (Tor Vergata, Rome)   and SuperKEKB \cite{SKB} at Tsukuba.


\begin{thebibliography}{00}
\bibitem{CC}
Inclusion of  charge conjugate modes is implied.
\end{thebibliography}
\end{document}